\documentclass[epj]{svjour}
\usepackage{graphics}
\begin{document}
\title{Associated strangeness production at threshold}
\author{ 
         P.~Kowina\inst{1,2}      \and
         M.~Wolke\inst{1}               \and
         H.-H.~Adam\inst{3}             \and
         A.~Budzanowski\inst{5}         \and
         R.~Czy{\.{z}}ykiewicz\inst{4}  \and
         D.~Grzonka\inst{1}             \and
         M.~Janusz\inst{4}              \and
         L.~Jarczyk\inst{4}             \and
         B.~Kamys\inst{4}               \and
         A.~Khoukaz\inst{3}             \and
         K.~Kilian\inst{1}              \and
         T.~Lister\inst{3}              \and
         P.~Moskal\inst{1,4}            \and
         W.~Oelert\inst{1}              \and
         T.~Ro{\.{z}}ek\inst{1,2}       \and
         R.~Santo\inst{3}               \and
         G.~Schepers\inst{1}            \and
         T.~Sefzick\inst{1}             \and
         M.~Siemaszko\inst{2}           \and
         J.~Smyrski\inst{4}             \and
         S.~Steltenkamp\inst{3}         \and
         A.~Strza{\l}kowski\inst{4}     \and
         P.~Winter\inst{1}              \and
         P.~W{\"u}stner\inst{6}         \and
         W.~Zipper\inst{2}               
}
%
%
\institute{IKP, Forschungszentrum J\"{u}lich, D-52425 J\"{u}lich, Germany
     \and Institute of Physics, University of Silesia, PL-40-007 Katowice, Poland
     \and IKP, Westf\"{a}lische Wilhelms--Universit\"{a}t, D-48149 M\"{u}nster, Germany
     \and Institute of Physics, Jagellonian University, PL-30-059 Cracow, Poland
     \and Institute of Nuclear Physics, PL-31-342 Cracow, Poland 
     \and ZEL,  Forschungszentrum J\"{u}lich, D-52425 J\"{u}lich,  Germany
}
\date{Received: date / Revised version: date}
%
\abstract{
The associated strangeness dissociation at threshold has been studied at the
COSY--11 facility measuring the hyperon -- and the  $ K^+ K^- $ meson pair
production.\\
Measurements of the near threshold $\Lambda$ and $\Sigma^0$ production
via the $pp \rightarrow p K^+ \Lambda / \Sigma^0$ reaction~\cite{sew99}
at COSY--11 have shown that the $\Lambda/\Sigma^0$ cross section 
ratio exceeds the value at high excess energies 
($\mbox{Q}~\ge$~300~MeV~\cite{bal88}) by an order of magnitude.
For a better understanding additional data have been taken
between 13~MeV and 60~MeV excess energy.  \\
The near threshold production of the charged kaon--antikaon pair is 
related to the discussion about the nature of the scalar states in the 
$1\,\mbox{GeV}/\mbox{c}^2$ mass range, i.e. the  $f_0\left(980\right)$
and $a_0\left(980\right)$~\cite{kre97}. The interpretation as a $K \overline{K}$ molecule is
strongly dependent on the  $K$-$\overline{K}$ interaction which can be studied via the
production channel.
A first total cross section value on the reaction $pp \rightarrow pp K^+ K^-$
at an excess energy of $17 \,\mbox{MeV}$~\cite{que01} i.e. below the $\phi$ 
production threshold was measured.
} 
%
\maketitle
%

\section{THE $\Lambda/\Sigma^0$ PRODUCTION RATIO CLOSE TO THRESHOLD}
\label{sec:1}
One of the main investigations of the COSY--11 collaboration is the associated 
strangeness production of the neutral $\Lambda$ and $\Sigma^0$ hyperons in the 
reactions $pp \rightarrow p K^+ \Lambda/\Sigma^0$. Since the quark structures 
of these hyperons are analogous to each other one can expect similar production 
mechanisms. 
In such a case the cross section ratio ${\cal{R}}_{\Lambda/\Sigma^0} 
\equiv \frac{\sigma(pp \to pK^+\Lambda)}{\sigma(pp \to pK^+\Sigma^0)}$ 
should be mainly determined by the isospin relation which leads  
to ${\cal{R}}_{\Lambda/\Sigma^0}=3$  what is in line with the ratio 
of about 2.5 observed at high excess energies 
($\mbox{Q} \ge 300\,\mbox{MeV}$)~\cite{bal88}. 
Very close to threshold, in the range of 
excess energies $\mbox{Q} \le 13\,\mbox{MeV}$, the total cross sections for the  
$\Lambda$ and $\Sigma^0$ 
hyperon production were measured  exclusively at the COSY--11 facility~\cite{bra96,wol00} at 
COSY J\"ulich~\cite{mai97}. The most remarkable feature of the data~\cite{sew99,bal98} 
was that at the same excess energy the total cross section for the $\Sigma^0$ production 
appeared to be about a factor of $28^{\,+6}_{\,-9}$ smaller than for the $\Lambda$ particle. 

        Enhancements in the missing mass distribution at the $\Lambda p$ and $\Sigma N$ 
thresholds observed in inclusive $K^+$ production data, taken at SPES~4~\cite{sie94}  
in proton-proton scattering at $\mbox{Q} = 252\,\mbox{MeV}$, both 
having about the same magnitude, suggest a strong $\Sigma N \rightarrow \Lambda p$ 
final state conversion. This conversion might be responsible for the decrease of the 
$\Sigma^0$ production yield close to threshold as seen in the COSY--11 data.
Strong $\Sigma N \rightarrow \Lambda p$ conversion effects are also suggested when 
interpreting the results of $K^-$ scattering on deu\-terons~\cite{tan69} where a sharp peak is 
clearly seen at an effective mass of the $\Lambda$-proton system  
m$_{\Lambda p} =2131$~MeV/c$^2$ corresponding exactly to the $\Sigma^0 p$ threshold. 

However, in calculations within the J\"ulich meson exchange model~\cite{gas00} 
the final state conversion is rather excluded as a dominant origin of the 
observed $\Sigma^0$ suppression. In these calculations both $\pi$ and $K$ exchange 
are taken into account with inclusion of the final state interaction (FSI) effects. 
$\Lambda$ production is found 
to be dominated by kaon exchange, what is  in line with the experimental results 
obtained by the DISTO collaboration~\cite{bal99} at higher excess 
energies ($\mbox{Q} = 430\,\mbox{MeV}$), where the importance of $K$ exchange 
is confirmed by a measurement of the polarisation transfer coefficient.
On the other hand,  in the case of $\Sigma^0$ production both $\pi$ and $K$ 
exchange are found to contribute with about the same strength. 
A destructive interference of the $\pi$ and $K$ exchange, suggested by Gasparian et 
al.~\cite{gas00}, is able to describe the suppression of the $\Sigma^0$ production 
observed in the close--to--threshold data.

Studies of the production ratio in~\cite{sib00} consider two different models:
either a $\pi$ plus $K$ exchange approach or the excitation of intermediate 
$N^*$ resonances via an exchange of $\pi$-- and heavier non--strange mesons, 
where the $N^*$'s  couple to the $K^+ Y$ channel~\cite{tsu}, however,  
any interference of the amplitudes are neglected.

The latter mechanism is also taken into account in an effective Lagrangian 
approach~\cite{shy01} where the strangeness production mechanism is modeled by the exchange 
of $\pi$, $\rho$, $\omega$ and $\sigma$ mesons, which excite the nucleon resonances 
$N^*(1650)$, $N^*(1710)$, and $N^*(1720)$. In both calculations experimental 
data are reproduced within a factor of two.

        The one-boson exchange calculation performed by La\-get~\cite{laget} taking into 
account interference effects of pion and kaon exchange only  by selecting the relative sign for
these two mechanism to maximise the cross section reproduce  not only the data of the 
$\Lambda/\Sigma^0$ ratio  within a factor of two but also  the polarisation transfer results of 
the DISTO experiment~\cite{bal99}.  

\begin{figure}
\resizebox{0.4\textwidth}{!}{%
  \includegraphics{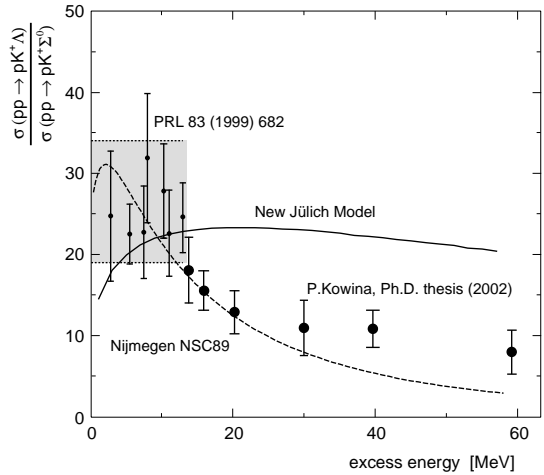}
}
\vspace{-0.1cm}
\caption{Energy dependence of the cross section ratio for
$\Lambda/\Sigma^0$ production in proton--proton collisions. 
Experimental data within the range up to $13~\mbox{MeV}$ are from~\cite{sew99}, 
data at higher excess energies from~\cite{kow02}. 
Calculations are performed within the J\"ulich meson exchange model, assuming a 
destructive interference of $K$ and $\pi$ exchange~\cite{gas02} and employing the 
microscopic $YN$ interaction models Nijmegen NSC89 (dashed line~\cite{mae89}) and the 
new J\"ulich model (solid line~\cite{hai01a}), respectively.
}
\label{fig:1}
\vspace{-0.35cm}
\end{figure}

        Recent COSY--11 measurements~\cite{kow02} extend the 
$\Lambda/\Sigma^0$ production ratio in proton-proton collisions up to an excess energy of 
$\mbox{Q} =$~60~MeV, what allows to study the behaviour of the cross section ratio in the 
transition region between the low energy range $\mbox{Q} \le$~13~MeV and data at high excess 
energies $Q \gg$~60~MeV. Together with the new~\cite{kow02} and 
earlier~\cite{sew99} experimental data calculations obtained within  the 
approach of Gasparian et al.~\cite{gas00} are presented in figure~\ref{fig:1}, where a 
destructive interference of $\pi$ and $K$ exchange is assumed, with different choices of the 
hyperon-nucleon interaction model for low--energy scattering in the final state used.
        The results of the calculations are very sensitive on the off-shell properties of the 
microscopic hyperon--nucleon interaction. 
 
Both the rather good description of the experimental data very close to threshold by the 
J\"ulich model~A~\cite{hol89} as well as the fair agreement for the  Nijmegen model 
(dashed line in Fig.~\ref{fig:1}) with the right tendency of the cross section ratio should
not regarded as being very conclusive. In the case of the Nijmegen model an explicit isospin 
symmetry breaking had to be introduced~\cite{mae89}. As a consequence the relation between  
amplitudes of the $\Sigma^{\pm}p$ and $\Sigma^0p$ channels is not uniquely 
defined~\cite{haidpriv}.

        As already emphasised in~\cite{gas02}, both constant elementary amplitudes
and only S--waves in the final state may not be justified for excess energies above 
$20\,\mbox{MeV}$  and thus the calculation based on the new J\"ulich model~\cite{hai01a} 
(solid line in fig.~\ref{fig:1}) does not reproduce the excitation function 
of the experimental cross section ratios.

         The data for the $\Lambda$ production in the excess energy range 
up to 60~MeV are described fairly well by the calculations of the phase space behaviour 
modified by the $p$-$\Lambda$ FSI~\cite{kow02} being in line with the scattering parameters 
from~\cite{Bale}. 
Contrary, in 
the case of $\Sigma^0$ there is almost no deviation from the phase space behaviour  in the 
energy dependence of the cross section for $\Sigma^0$ production, which might indicate a very 
weak $p$-$\Sigma^0$ FSI~\cite{kow02}. However, it should be noted that the apparently weak 
influence of the $p$-$\Sigma^0$ FSI could be feigned by either higher partial wave 
contributions or an energy dependence of the elementary amplitude~\cite{gas02}.
Therefore further  measurements at an excess energy of $Q \approx 60~\mbox{MeV}$ are 
highly desirable to study  the angular distribution of the produced $\Lambda$ and $\Sigma^0$ 
hyperons.

\section{EXCLUSIVE KAON--ANTIKAON PRODUCTION AT COSY--11}
\label{sec:2}
        Different interpretations of the structure of the scalar resonances
$f_0(980)$ and $a_0(980)$ are known~\cite{thom02,Mor93+}. Some motivations for measurements
of the  $K^+ K^-$ production were calculations within the J\"ulich meson exchange model
for the $\pi \pi$ and $\pi \eta$ scattering. The results of these calculation are very 
sensitive on a strength of the  $K \overline{K}$ interaction~\cite{kre97}.   
Therefore, measurements of 
the energy dependence of the cross section can help to confirm or exclude the thesis, that 
the production of $K^+ K^-$ leads via excitation of the intermediate resonance. 
        Unfortunately those calculations are done only for  $\pi \pi$ scattering,
however, similar effects are expected in the case of the $p p$ 
interaction~\cite{hai01b}.

\begin{figure}
\resizebox{0.4\textwidth}{!}{%
  \includegraphics{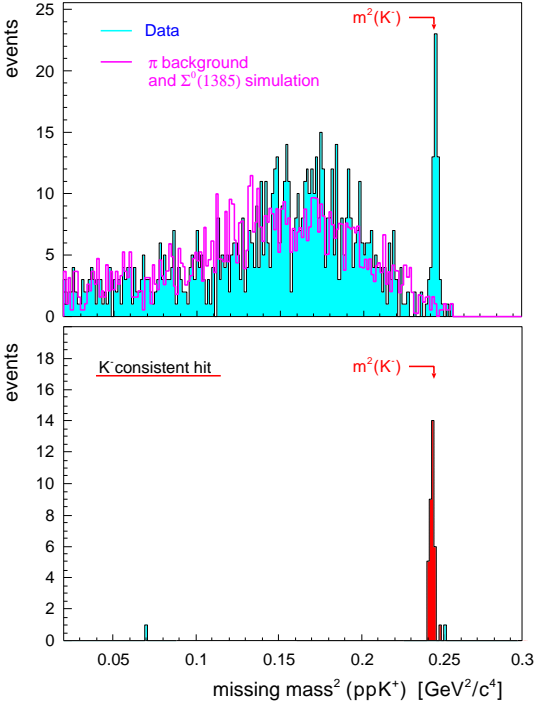}
}
\vspace{-0.1cm}
\caption{Missing mass distribution with respect to an identified 
($\mbox{p}\mbox{p}\mbox{K}^+$) subsystem at an excess energy of 
$17\,\mbox{MeV}$ above the $\mbox{p}\mbox{p} \rightarrow 
\mbox{p}\mbox{p}\mbox{K}^+\mbox{K}^-$ production threshold 
without (a) and with (b) $\mbox{K}^-$ detection~\cite{que01}.} 
\label{fig:2}       
\vspace{-0.6cm}
\end{figure}

        From the reconstruction of the full four--momentum vectors for all positively charged 
ejectiles one obtains the missing mass spectrum  of the $(pp K^+)$ system shown in the upper 
part of figure~\ref{fig:2} where a clear peak with a resolution (FWHM) of 
$\approx 2\,\mbox{MeV}/\mbox{c}^2$ is seen at the mass of the charged
kaon. The physical background seen is mainly due to the excitation of the 
hyperon resonances $\Lambda\left(1405\right)$ and $\Sigma\left(1385\right)$, where the proton 
originating from the hyperon resonance decay is detected.

        Requiring an additional $K^-$ hit in the dedicated  negative particle detector 
installed at the COSY--11 facility~\cite{bra96} one obtains an almost background free spectrum 
of the missing mass of the $p p K^+$ system shown in the lower part of the Fig.~\ref{fig:2}. 
The number of entries in the $K^-$ peak is slightly reduced compared to the upper figure due 
to the influence of the kaon decay and acceptance. The analysis resulted in a first total 
cross section for the elementary $K^+ K^-$ production below the $\Phi$ threshold at 
$Q = 17\,\mbox{MeV}$ measured in the proton--proton scattering which is 
$\sigma = 1.80 \pm 0.27^{+0.28}_{-0.35}\,\mbox{nb}$ with statistical and systematical errors, 
respectively~\cite{que01}. The cross section for the 
$p p \to p K^+ \Lambda$~\cite{sew99,bal98,kow02} reaction which is the elementary $K^+$ 
production is two orders of magnitude higher compared to the cross section for the elementary 
$K^-$ production in the $p p \to p p K^+ K^-$ reaction at the corresponding excess energies.

        At the present stage it is not possible to judge whether $K^+ K^-$ proceeds via a 
resonant production with the excitation of the $f_0(980)$ and $a_0(980)$ scalar resonances.

        The energy dependence of the total cross section for $K^+ K^-$ -- 
below~\cite{que01} and above~\cite{bal01} the $\Phi$ threshold might be compared to
data for $\eta^\prime$~\cite{bal88,etap} production, where for an excess energy range 
$100 \le \mbox{Q} \le 1000\,\mbox{MeV}$ the excitation function is well 
described by a three--body phase space ($\sigma \propto \mbox{Q}^2$). To describe the data 
below $100\,\mbox{MeV}$ at least the FSI between the final state protons and possibly 
even the FSI between the final state proton and meson have to be considered.

        Not so for the $K^+ K^-$ production, where calculations based on a one--boson 
exchange~\cite{sib97} neglecting FSI effects give significantly different results than simply 
assuming a four--body phase space behaviour. Contrary to the  $\pi N \rightarrow \eta^\prime N$ 
amplitudes the $K^+ p$ and especially the $K^- p$ amplitudes are strongly energy 
dependent~\cite{sib01}.
The reason might be a compensation of the interaction of the two strongly interacting 
subsystems  $pp$ and $K^- p$ in the final state or an additional degree 
of freedom given by the four--body exit channel. In such a case the influence of the FSI effects 
should be more pronounced at the $K^+ K^-$ production threshold~\cite{sib01}.

        Additional data were taken at the COSY--11 facility at excess energies
$10\,\mbox{MeV}$ and $28\,\mbox{MeV}$, i.e.\ close to the $K^+ K^-$ production
threshold and slightly below threshold for the $\Phi$ production. The data analysis is presently
in progress.


\begin{thebibliography}{}
%
%
%
%
\bibitem{sew99} S.\ Sewerin et al., Phys.\ Rev.\ Lett.~\textbf{83} (1999) 682. 
\bibitem{bal88} A. Baldini et al., \textit{\em Total Cross--Sections for 
        Reactions of High--Energy Particles}, (Landolt--B\"ornstein, 
        {\em New Series} I/12, Springer, Berlin 1988).
\bibitem{kre97} O.\ Krehl, R.\ Rapp, J.\ Speth, Phys.\ Lett.\ \textbf{B~390} (1997) 23.
\bibitem{que01} C. Quentmeier et al., Phys. Lett. \textbf{B~515} (2001) 276.
\bibitem{bra96} S.\ Brauksiepe et al., Nucl.\ Instr. \& Meth.~\textbf{A~376} (1996) 397.
\bibitem{wol00} M.\ Wolke, Nucl.\ Phys.\ News \textbf{9} (1999) 27.
\bibitem{mai97} R.\ Maier, Nucl.\ Instr. \& Meth.~\textbf{A~390} (1997) 1
\bibitem{bal98} J.T.\ Balewski et al., Phys.\ Lett.\ \textbf{B~420} (1998) 211.
\bibitem{sie94} R.\ Siebert et al., Nucl.\ Phys.\ \textbf{A~567} (1994) 819.
\bibitem{tan69} T.H.\ Tan, Phys.\ Rev.\ Lett. \textbf{23} (1969) 395.
\bibitem{gas00} A.\ Gasparian et al., Phys.\ Lett.\ \textbf{B~480} (2000) 273. 
\bibitem{bal99} F.\ Balestra et al., Phys.\ Rev.\ Lett. \textbf{83} (1999) 1534.
\bibitem{sib00} A.\ Sibirtsev et al., e--Print nucl--th/0004022 (2000).
\bibitem{tsu} K. Tsushima et al., Phys.\ Rev.~\textbf{C 59} (1999) 369.
\bibitem{shy01} R. Shyam et al., Phys. Rev. \textbf{C~63} (2001) 022202.
\bibitem{laget} J.--M. Laget, Phys.\ Lett.~\textbf{B 259} (1991) 24, 
        Nucl.\ Phys.~\textbf{A 691} (2001) 11c.
\bibitem{kow02} P. Kowina, Ph.D. thesis, Silesian Univ. Katowice (2002).
\bibitem{gas02} A. Gasparyan, FZ--J\"ulich, 
        Matter and Materials \textbf{11} (2002) 205. 
\bibitem{hol89} B. Holzenkamp et al., Nucl. Phys. \textbf{A~500} (1989) 485.
\bibitem{mae89} P.M.M. Maessen et al., Phys. Rev. \textbf{C~40} (1989) 2226.
\bibitem{haidpriv} J. Haidenbauer, private communications. 
\bibitem{hai01a} J. Haidenbauer et al., AIP Conf. Proc. \textbf{603} (2001) 421.
\bibitem{Bale} J.T.\ Balewski et al., Eur.\ Phys. \ J \textbf{A~2} (1998) 99.
\bibitem{thom02} U.Thoma, proceedings of this conference. 
\bibitem{Mor93+} D. Morgan, M.R. Pennington, Phys. Rev. \textbf{D~48} (1993) 1185; 
        F. Kleefeld et al., Phys. Rev. \textbf{D~66} (2002) 034007; 
        R.L. Jaffe, Phys. Rev. \textbf{D~15} (1977) 267; 
        F.E. Close, Rep. Prog. Phys. \textbf{51} (1988) 833;
        F.E. Close et al., Phys. Lett. \textbf{B~319} (1993) 291;
        J. Weinstein, N. Isgur, Phys. Rev. \textbf{D~41} (1990) 2236;
        D. Lohse et al., Nucl. Phys. \textbf{A~516} (1990) 513;
        Z.S. Wang et al., Nucl. Phys. \textbf{A~684} (2000) 429c.
\bibitem{hai01b} J. Haidenbauer, FZ--J\"ulich, 
        Matter and Materials \textbf{11} (2002) 225.
\bibitem{etap} F. Hibou et al., Phys.\ Lett.~\textbf{B~438} (1998) 41;
        P. Moskal et al., Phys.\ Rev.\ Lett.~\textbf{80} (1998) 3202;
        P. Moskal et al., Phys.\ Lett.~\textbf{B 474} (2000) 416;
        F. Balestra et al., Phys.\ Lett.~\textbf{B 491} (2000) 29.
\bibitem{bal01} F. Balestra et al., Phys. Rev. \textbf{C 63} (2001) 024004.
\bibitem{sib97} A. Sibirtsev et al., Z. Phys. \textbf{A~358} (1997) 101.
\bibitem{sib01} A. Sibirtsev, FZ--J\"ulich, Matter and Materials \textbf{11} (2002) 239. 
\end{thebibliography}
\end{document}